\documentclass[%
 reprint,
superscriptaddress,
 amsmath,amssymb,
 aps,
]{revtex4-1}
\usepackage[inline]{enumitem}
\usepackage{graphicx}

\usepackage{float}
\usepackage{xcolor}
\usepackage{multirow}

\begin{document}
	
	\title{Charge density and redox potential of LiNiO$_2$ using ab initio diffusion quantum Monte Carlo}
	\author{Kayahan Saritas}
	\altaffiliation[Current affiliation: ]{Applied Physics, Yale University, New Haven, Connecticut 06520, USA}
	\affiliation{Materials Science and Engineering Department, Massachusetts Institute of Technology, Cambridge, Massachusetts 02139,  USA}
	\author{Eric R. Fadel}
	\affiliation{Materials Science and Engineering Department, Massachusetts Institute of Technology, Cambridge, Massachusetts 02139,  USA}
	\affiliation{John A. Paulson School of Engineering and Applied Sciences, Harvard University, Cambridge, Massachusetts 02138, United States}
	\affiliation{Robert Bosch LLC, Research and Technology Center North America, 255 Main St, Cambridge, Massachusetts 02142, USA}
	\thanks{K. Saritas and E. R. Fadel contributed equally to this work}
	\author{Boris Kozinsky}
	\affiliation{John A. Paulson School of Engineering and Applied Sciences, Harvard University, Cambridge, Massachusetts 02138, United States}
	\affiliation{Robert Bosch LLC, Research and Technology Center North America, 255 Main St, Cambridge, Massachusetts 02142, USA}
	\author{Jeffrey C. Grossman}
	\affiliation{Materials Science and Engineering Department, Massachusetts Institute of Technology, Cambridge, Massachusetts 02139,  USA}
	
	
	
	
	\begin{abstract}
	Electronic structure of layered LiNiO$_2$ has been controversial despite numerous theoretical and experimental reports regarding its nature. We investigate the charge densities, lithium intercalation potentials and Li diffusion barrier energies of Li$_x$NiO$_2$ ($0.0 < x < 1.0$) system using a truly \textit{ab-initio} method, diffusion quantum Monte Carlo (DMC). We compare the charge densities from DMC and density functional theory (DFT) and show that local and semi-local DFT functionals yield spin polarization densities with incorrect sign on the oxygen atoms. SCAN functional and Hubbard-$U$ correction improves the polarization density around Ni and O atoms, resulting in smaller deviations from the DMC densities.  DMC accurately captures the p-d hybridization between the Ni-O atoms, yielding accurate lithium intercalation voltages, polarization densities and reaction barriers. 
	\end{abstract}
	
	
	\maketitle
	
	\section{Introduction}
	Lithium-ion battery technologies have undergone tremendous advances leading to major developments in a surge of applications from mobile technologies to electric vehicles \cite{Thackeray1983,Motavalli2015,Mizushima1981}. However, further improvements in storage density are still needed. Developing cathode materials that are suitable for reversible energy storage is a challenging task which requires multiscale materials discovery approach. Many cathode materials have been discovered and studied using experimental methods \cite{Parz2013,Iwaya2013,Motohashi2009,Takahashi2007,Godinez1996,Hertz2008}. However, these efforts can be accelerated using atomic scale theoretical and computational approaches \cite{Urban2016,Seo2015,VanderVen1998,Shao-Horn2003,Wolverton1998} that yield high accuracy redox potentials. 
	
	Density functional theory (DFT) \cite{Kohn1965, Hohenberg1964} is often used to predict redox potentials, band gaps and the formation energies of transition metal oxides \cite{Ceder1999, Ong2011, Urban2016, Marianetti2004, Jain2013} due to its favorable balance of computational cost and accuracy. Li-ion battery cathode materials are based on redox-active transition metal oxides, fluorides, phosphates etc. In these materials, local or semi-local DFT exhibits non-systematic errors because of significant self-interaction errors from the localized \textit{d} electrons. To correct this self interaction error, it is common to apply an impurity model (e.g. DFT with Hubbard model correction referred to as DFT+$U$ \cite{Anisimov1991, Dudarev1998}) or to include some portion of the exact exchange \cite{Heyd2003} (hybrid-DFT). These methods involve adjustable parameters that are often tuned to increase the accuracy of various properties, including redox potentials. Transferability of these parameters across the family of transition metal oxides (e.g. nickelates, cobaltites) is questionable, thus the \textit{ab-initio} character of the calculations are reduced in favor of increased accuracy. Therefore, it is difficult to understand and design electronic and energetic properties of cathode materials using available DFT-based methods when no experimental guidance is available \cite{Tarascon2001IssuesAC,whittingham2004lithium,bhatt2015recent}. 

	Using a judicious choice of the $U$ value on the transition metal atom can help yield reasonable redox potentials.
	$U$ values can be determined self consistently using linear response theory \cite{cococcioni2005linear, himmetoglu2014hubbard, Kulik2006}. However, the $U$ values determined through linear response can depend on the material and valence which changes during delithiation \cite{Wang2006,Wang2007, Hautier2012, Seo2015, Delmas1999, Marianetti2004}.
	Due to the strong electronegativity of oxygen, valence electrons on the transition metal species are rearranged upon changing the Li concentration, which formally necessitates a separate $U$ value at every state of charge for the the transition metal species. This is crucial to the accuracy of redox potentials since the average redox potential for Li extraction is calculated using the following equation \cite{mckinnon1983physical,aydinol1997ab}:
    	
    \begin{equation}
    V=-\frac{        G[Li_{x2}] - G[Li_{x1}]    -(x_2-x_1)G[Li]      } {x_2-x_1}
    \label{eqn:intercalation}
    \end{equation} where G is the Gibbs free energy of the compounds at Li concentrations of x$_1$ and x$_2$. Typically DFT (or DMC) ground state energies can be used to replace the Gibbs free energies with very little error. Therefore, one would need to perform three calculations to determine the average voltage at different Li concentrations: $Li_{x1}$, $Li_{x2}$ and metallic Li. However, it is not clear whether DFT energies with different $U$ values can be used for calculating energy differences corresponding to the redox voltages. 

	In this work, we aim to eliminate most of the mentioned challenges and calculate electronic and energetic properties of Li$_x$NiO$_2$ using a fundamentally different approach: diffusion quantum Monte Carlo (DMC)\cite{Foulkes2001, Needs2010, Shulenburger2013}. DMC is a many-body method which has been successfully used to calculate equilibrium geometries, defect and crystalline formation energies, exchange coupling constants and quasiparticle gaps of transition metal oxides with near chemical accuracy, comparable to the coupled cluster calculations in quantum chemistry applications \cite{Saritas2017, Kolorenc2010, Yu2017, Yu2015, Santana2015, Luo2016, Kylanpaa2017, Foyevtsova2014, Needs2002, Williamson1998, Mitra2015, Schiller2015, Saritas2018a, Saritas2018b, Saritas2019, Saritas2017f}. Using DMC, we study the Li intercalation voltages, charge density distributions and Li diffusion barriers of Li$_x$NiO$_2$, where $0 < x < 1$. We then benchmark our DMC calculations with DFT and DFT+$U$. We highlight important differences between these methods and  demonstrate the limitations of DFT+$U$ corrections.
	
    \begin{figure}
	\includegraphics[width=0.5\textwidth]{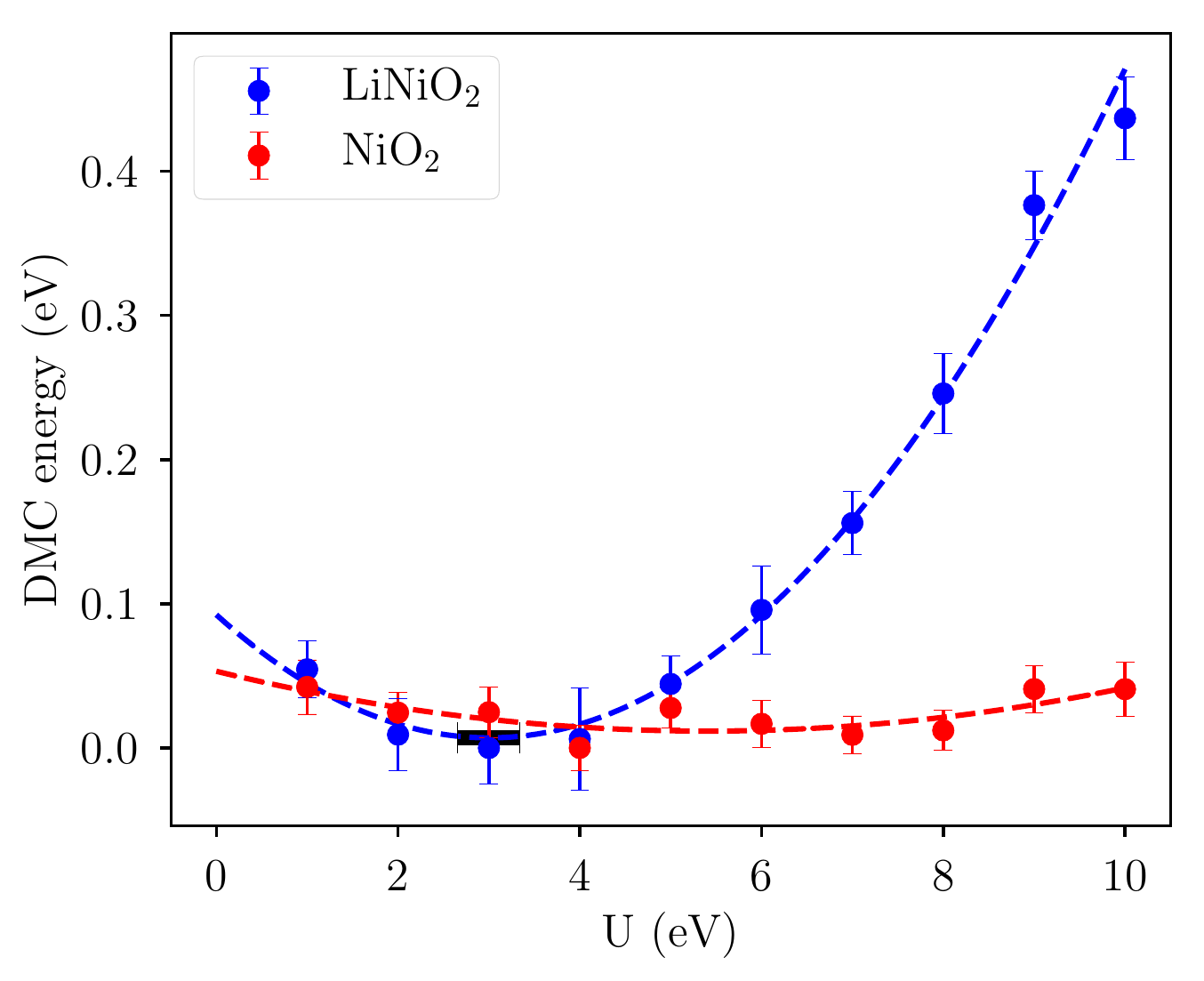}
	\caption{(Color online) DMC wavefunction optimization for NiO$_2$ and LiNiO$_2$ structures using +$U$ interactions (within LDA) on the \textit{d}-shell of Ni atoms. Minimum DMC energies for both structures are set to zero for better visual comparison.}
	\label{fgr:plusu}
    \end{figure}	

\begin{figure}
	\includegraphics[width=0.5\textwidth]{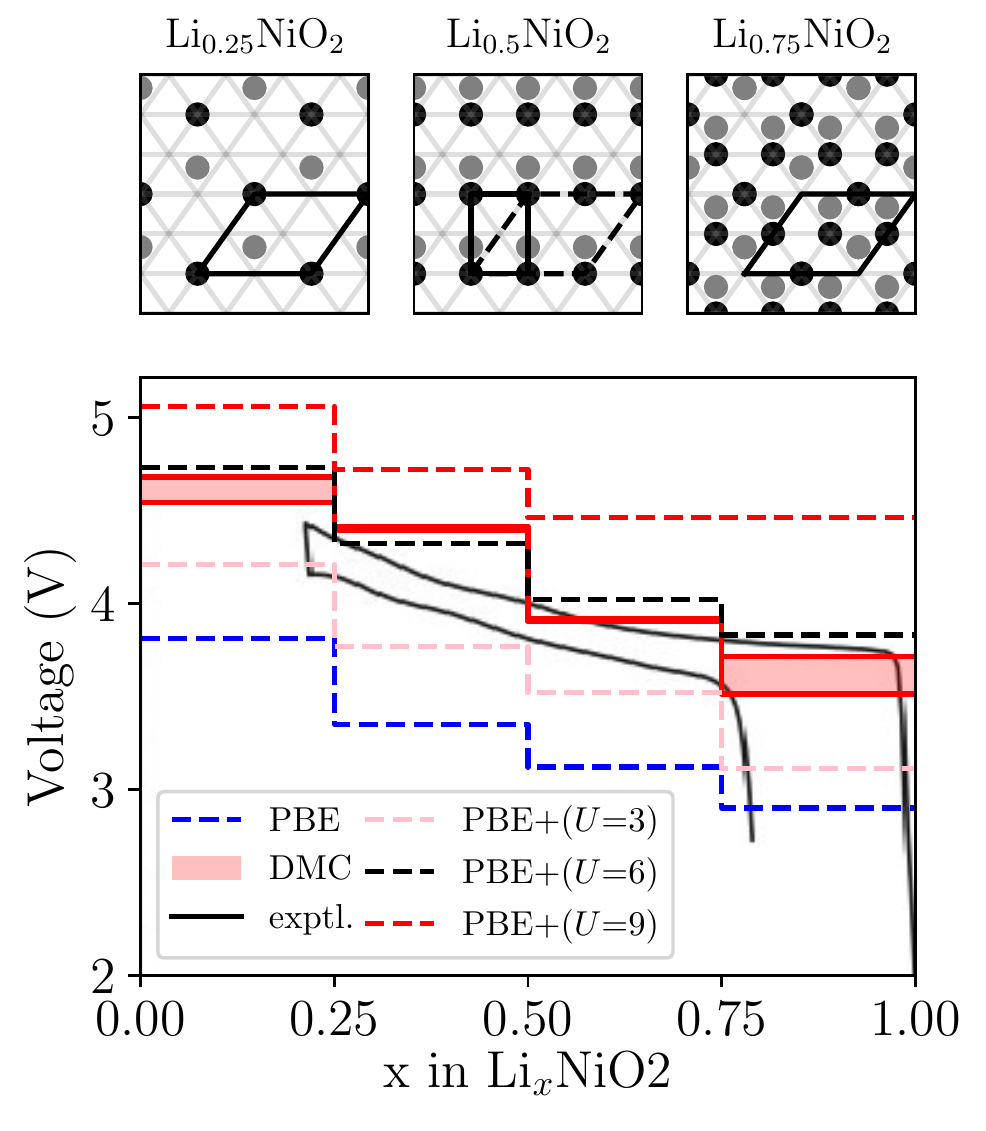}
	\caption{(Color online) Upper three figures indicate the Li-vacancy ordering in partially lithiated structures from on the (001) plane. Gray and black circles denote Li atoms on different planes. Black colored Li atoms are on the (001) surface, and they are separated from the gray colored Li atoms with a Ni-O layer. Black lines show the primitive cell boundaries, while the dashed lines in Li$_{0.5}$NiO$_2$ indicate the 14-atom supercell that is used in our DMC calculations which has the similar texture with Li$_{0.25}$NiO$_2$ and Li$_{0.75}$NiO$_2$. In the lower figure, Li intercalation voltages of Li$_x$NiO$_2$ are shown using DFT+$U$ and DMC calculations. Shaded areas of the DMC curves show the error bars of the voltages obtained. Computational voltages are overlayed on the experimental voltage curve with its measured hysteresis \cite{Delmas1980}}
	\label{fgr:volt}
\end{figure}
\section{Methods}\label{sec:methods}
In this work we used DFT and DMC ground-state energies to determine the redox potential of LiNiO$_2$ as a function of Li concentration.
We used Dudarev's Hubbard-$U$ \cite{Anisimov1997, Dudarev1998} corrected PBE \cite{Perdew1996} DFT functional to benchmark our DMC results. For DFT calculations, all geometries are optimized separately for each functional. We used  Vienna ab-initio simulation package\cite{Kresse1994,Kresse1996a} (\textsc{VASP}) code for all the reported DFT energies and redox potentials. In these calculations, we used projector-augmented wave pseudopotentials \cite{Kresse1999} and a kinetic energy of 520 eV.

Calculating DMC ground-state energies of involves multiple practical steps. Here, we broadly explain main steps involved in a general way and in the following paragraph we will discuss the technical details. First, a trial (guiding) wavefunction must be generated often using single particle Slater determinants (DFT, DFT+$U$, hybrid-DFT) \cite{Schiller2015} or using configuration-interaction (CI) methods \cite{Dash2018, Scemama2018}. The trial wavefunction defines the nodal surface of the fixed-node DMC calculation where the wavefunction goes to zero. However, the CI methods are largely limited to finite systems as they are computationally rather demanding. Therefore, DFT based methods have been largely utilized in periodic systems. In this approach the nodal surface is often optimized by either varying the $U$ interaction parameter or the exact exchange ratio in the hybrid-DFT functionals \cite{Yu2017, Doblhoff-Dier2016, Shin2017, Kylanpaa2017}. Second, Jastrow parameters (correlation functions) are added to the guiding wavefunction and then optimized to further capture many-body correlations in the system. Jastrow parameters are defined for many-body interactions such as electron-electron and electron-electron-ion as examples. Optimizing these parameters require evaluation of expensive stochastic derivatives, hence they are performed using variational Monte Carlo (VMC). Finally, DMC calculations are performed using the trial wavefunction and the optimized Jastrow parameters. The DMC calculations involve equilibration and statistics accumulation steps. We report the DMC energies obtained in the statistics accumulation step. Detailed information regarding the DMC method can be found in the literature and the Supplementary Information \cite{Foulkes2001, Needs2010, Shulenburger2013, Supp}. 

DMC and VMC calculations were performed using \textsc{QMCPACK}\cite{QMCPACK}, while DFT-VMC-DMC calculation workflows are generated using the \textsc{Nexus} \cite{Nexus} software suite.  We used PBE+$U$ functional to generate spin-up and spin-down trial wavefunctions using \textsc{Quantum Espresso}\cite{QE-2009} (QE) code. For Ni and O atoms, we used the norm-conserving RRKJ type pseudopotentials \cite{Krogel2016, OPIUM}, while for the Li atom we use BFD pseudopotentials\cite{Burkatzki2007} converted to Kleinman-Bylander form. These pseudopotentials are specifically constructed for DMC calculations and require very large kinetic energy cutoffs (350 Ry), hence they are not practical for DFT calculations. Therefore, in this work, QE code and the RRKJ pseudopotentials were only used towards the DMC calculations. The pseudopotentials in this work were well validated on similar systems using DMC, such as Li intercalation of multilayered graphene \cite{Ganesh2014}, formation energies, ground and excited states of NiO \cite{Mitra2015, Shin2017}.  
Each wavefunction we used in the DMC calculations were made of a Slater determinant and Jastrow factors, where both were optimized separately. Varying the $U$ interaction energy in PBE+$U$, we optimized the nodal surface of the trial wavefunction.  DFT+$U$ calculations yield monotonously increasing energies with increased U value. However, the $U$ value is used as a variational parameter in DMC, because with a fixed set of Jastrow parameters, the guiding wavefunction that has the nodal surface closest to exact nodal surface yields the exact ground state energy of that system \cite{Foulkes2001}. For NiO$_2$ and LiNiO$_2$, we performed this step in Fig. \ref{fgr:plusu}. Fig. \ref{fgr:plusu} will be further explained later. Jastrow parameters were optimized using subsequent VMC variance and energy minimization calculations using the linear method \cite{Umrigar2007} as implemented in \textsc{QMCPACK}. Cost function of the energy minimization is split as 95/5 energy and variance minimization, which is shown to provide a good balance for improvements in DMC with the resulting variance \cite{Umrigar2005}.

The DMC calculations are performed using a uniform gamma-centered 3x3x3 reciprocal grid on all the supercells studied, with a time step of 0.01 Ha$^{-1}$. We used the model periodic coulomb (MPC) interaction to eliminate spurious two-body interactions \cite{Drummond2008, Fraser1996}. The locality approximation \cite{Mitas1991} has been shown to yield smaller localization errors in Ni atoms \cite{Dzubak2017a}, compared to T-moves \cite{Casula2006}. Therefore the locality approximation was used throughout this work. 

In Fig. \ref{fgr:plusu}, the trial wavefunction optimization is shown for 12 and 16-atom cells in NiO$_2$ and LiNiO$_2$ using a 2x2x2 reciprocal grid with PBE+$U$. Here, the $U$ value is used as a variational parameter to optimize the nodal surface of the trial wavefunction. In practice, one can use different flavors of DFT functionals (local, semilocal, meta-GGA). However, it has been shown that optimized LDA+$U$ and PBE+$U$ trial wavefunctions yield identical DMC energies in NiO, while PBE+$U$ has smaller curvature in the DMC energy versus $U$ interaction energy curves \cite{Shin2017}. In Fig. \ref{fgr:plusu}, we find that the DMC total energy of LiNiO$_2$ is more sensitive to the $U$ value (a sharp minimum at $U=3.0$ eV, while for NiO$_2$ the DMC energies are statistically identical within the range of $U$ values studied. Therefore, we used PBE+$U=3.0$ eV on all the Ni atoms to generate DMC trial wavefunctions in this work. 

Geometry optimization in DMC is computationally demanding. Therefore, we fully optimized geometric structures using DFT-PBE. It has been shown PBE and PBE+$U$ can perform well against the experimental lattice parameters \cite{Chakraborty2018}.  For LiNiO$_2$, we used the \textit{R-3m} symmetry cell as our starting geometry for the geometry relaxation calculations\cite{Hirano1995}. We used 4x4 supercells in the \textit{xy}-plane to determine the minimum energy Li-vacancy ordering for the Li$_x$NiO$_2$ structures. Minimum energy structures for each Li concentration are then studied with DMC. The primitive cell lattice parameter these structures are shown in Table 1 at Supplementary Information \cite{Supp}. The (001) projections of the vacancy ordered structures we used in this work are shown in Fig. \ref{fgr:volt}, agreeing with theoretical and experimental findings \cite{VanderVen1998,reimers1992electrochemical,y2002first,arai1995characterization,Delmas1980,Delmas1999}. 

One and two body finite size effects must be controlled in periodic DMC calculations. One-body effects are controlled using twist averaging and using a twist averaging correction scheme similar to that proposed by Rajagopal et. al. \cite{Rajagopal1994} and as implemented in Ref. \citenum{Saritas2017}. Whereas the two-body effects require extrapolation to the infinitely large system. If the supercell size is very large, then the two-body effects become minimal and independent of the shape of the supercell. However, for a given volume the supercell that can be used in the DMC calculations is not unique. Therefore, we used \textit{optimal\_tilematrix} method in \textsc{Nexus}, to generate the supercell tiling matrices. \textit{Optimal\_tilematrix} method maximizes the minimum inscribing radius to reduce image interactions in all directions. However, the supercells with similar systems and lattice parameters can benefit from systematic error cancellations: e.g. achieve faster convergence on the energy differences between the Li$_x$NiO$_2$ supercells.
Therefore, for the 14 atom cell of Li$_{0.5}$NiO$_2$ we used a supercell (shown in Fig. \ref{fgr:volt}) which has similar lattice parameters to Li$_{0.25}$NiO$_2$ and Li$_{0.75}$NiO$_2$. 
An extrapolation scheme was used on the DMC charge densities to eliminate the bias from using a mixed estimator at the DMC level.
In DMC the charge density estimator does not commute with the fixed node DMC Hamiltonian \cite{Foulkes2001}. 
Hence, the collected DMC density is a mixed estimator between the pure fixed-node DMC and VMC densities. 
In order to obtain the pure fixed-node DMC density, the following extrapolation formulas can be used \cite{Foulkes2001}:
\begin{equation}
\rho_1 = 2\rho_{DMC}-\rho_{VMC}+\mathcal{O}[(\Phi-\Psi_T)^2] 
\label{eqn:extrapolation1}
\end{equation}
\begin{equation}
\rho_2 = 2\rho_{DMC}^2/\rho_{VMC}+\mathcal{O}[(\Phi-\Psi_T)^2]
\label{eqn:extrapolation2}
\end{equation}
where $\rho_{DMC}$ and $\rho_{VMC}$ are DMC and VMC charge densities respectively.
The accuracy of the estimators increases with the increased quality of the trial wavefunction, $(\Phi-\Psi_T)$, where $\Psi$ is the wavefunction from DMC Hamiltonian and the $\Psi_T$ is the trial wavefunction. Ideally in homogeneous systems both estimators above should yield identical results. However, in our case the pseudo Li$^{+}$ atoms donate almost all of their electrons, making the second extrapolation scheme in Eqn. \ref{eqn:extrapolation2} numerically challenging. This is because the VMC density in the denominator approaches zero near the charge depleted regions of pseudo Li$^{+}$ atoms. Therefore, the extrapolation scheme in Eqn. \ref{eqn:extrapolation1} was used throughout this work.

\section{Results and Discussion}\label{sec:res}
\subsection{Ground state calculations}\label{sec:ground}
In LiNiO$_2$, Ni is in an octahedral environment with a formal charge of 3+. Theoretical calculations predict that Ni$^{3+}$ has the $d^7$ ($t_{2g}^6e_g^{1}$) electronic configuration which yields a net magnetization of 1 $\mu$ on each Ni atom \cite{y2002first, ArroyoydeDompablo2001, ArroyoydeDompablo2003}.  However, it has been challenging to experimentally observe the long-range ferromagnetic ordering in LiNiO$_2$ \cite{Chen2011,kemp1990magnetic,reynaud2001orbital}. Plane-wave ab-initio calculations with periodic boundary conditions were used to find ferromagnetic ordering in the ground state \cite{zhou2004first,Chen2011}. In the layered NiO$_2$ structure, the Ni atom donates its $e_g$ electron to O atoms and formally becomes diamagnetic \cite{zhou2004first}. Our PBE+$U$ calculations agree with these findings, yielding a distribution of Ni$^{3+}$ and Ni$^{4+}$ atoms in  all Li$_x$NiO$_2$ structures based on the projection of the charge density on atomic orbitals.  Structural parameters and the total magnetic moments of the structures we studied are listed in the Supplementary Information \cite{Supp}. 

In Fig. \ref{fgr:volt} we show the Li-intercalation potential of the Li$_x$NiO$_2$ structures ($x=\{0.0, 0.25, 0.5, 0.75, 1.0\}$) , calculated using DMC and DFT methods. In the figure, theoretical results are overlayed on the experimental results from ref. \citenum{Delmas1980}. In Fig. \ref{fgr:volt}, we first show the primitive cells for each delithiated structures. DMC intercalation voltages show  excellent agreement with the experimental curve except for the slight overestimation between Li$_{0.25}$NiO$_2$ and Li$_{0.5}$NiO$_2$. From PBE+$U$=0 to 9 eV, the average voltage increases monotonically. All DFT functionals in Fig. \ref{fgr:volt}, except PBE+$U$=9 have the same step features as the voltage profile, while the step at $x$=0.75 disappears for PBE+$U$=9. Similar loss of stepwise features in the voltage curves has also been observed near complete Li intercalation with the HSE functional when the exact exchange ratio is increased from 0.17 to 0.25 \cite{Seo2015} in LiCoO$_2$. It has been suggested that at the high and low Li intercalation limits, different amounts of exact exchange would be required to reproduce the experiments \cite{Seo2015}. Our findings, in terms of the loss of stepwise features in LiNiO$_2$, are similar to LiCoO$_2$ \cite{Seo2015}. Hence, we demonstrate the challenges of using a constant $U$ vaue or exact exchange ratio in hybrid DFT functionals to reproduce the redox potentials across the Li intercalation limits. 

\begin{figure}
	\includegraphics[width=0.5\textwidth]{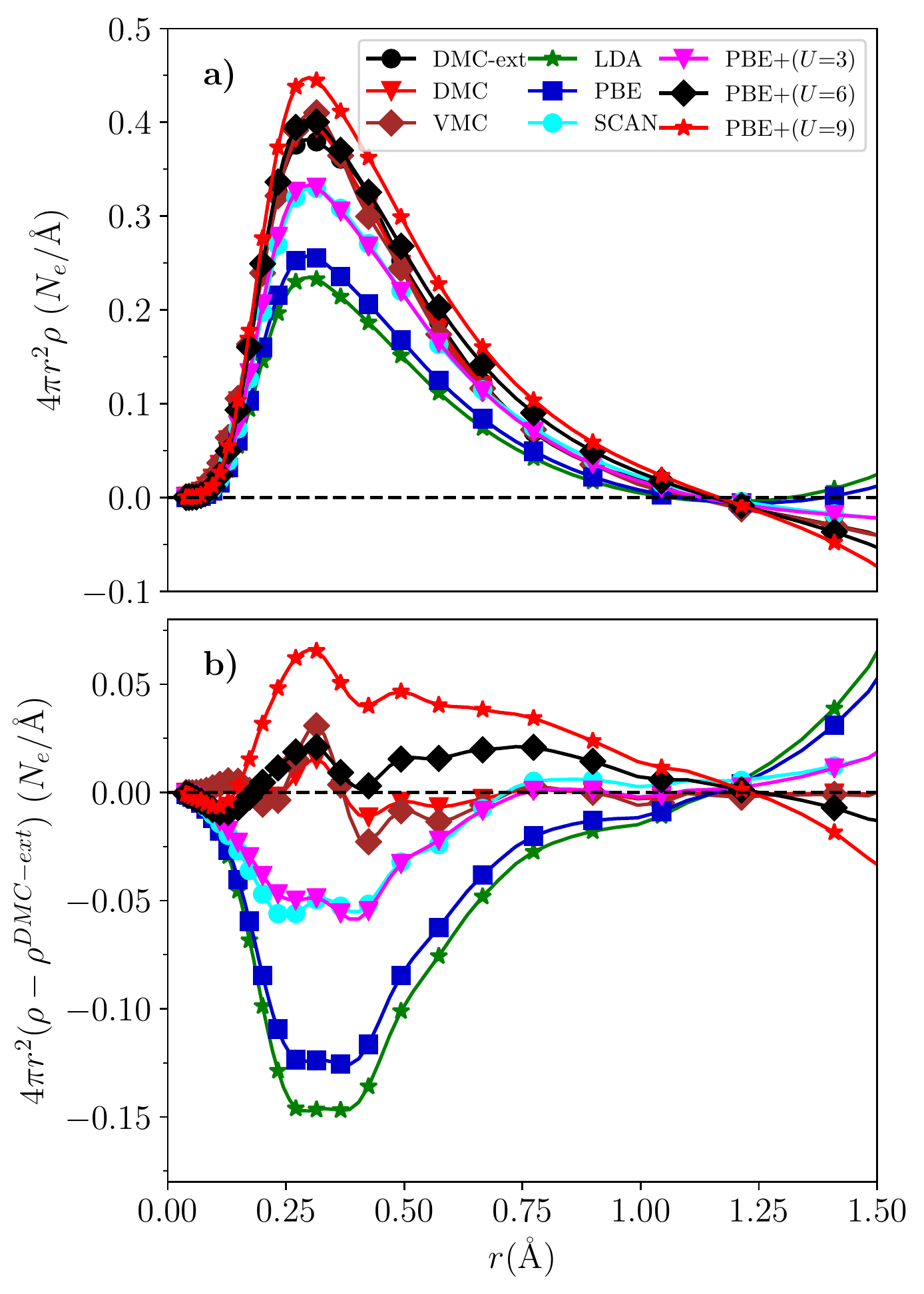}
	\caption{(Color online) a) Radial spin polarization density ($\rho=\rho_{\uparrow}-\rho_{\downarrow}$) and b) radial spin polarization density difference from extrapolated DMC around the Ni atom with various theoretical methods using RRKJ pseudopotentials. A $U$ value of 6 eV with PBE is the most accurate DFT charge density among the tested DFT functionals. PBE+$U$=6 also has the most accurate voltages in Fig. \ref{fgr:ni_density}. Every two out of three markers are omitted for clarity.}
	\label{fgr:ni_density}
\end{figure}

\subsection{Charge densities}\label{sec:chg}

\begin{figure}
	\includegraphics[width=0.5\textwidth]{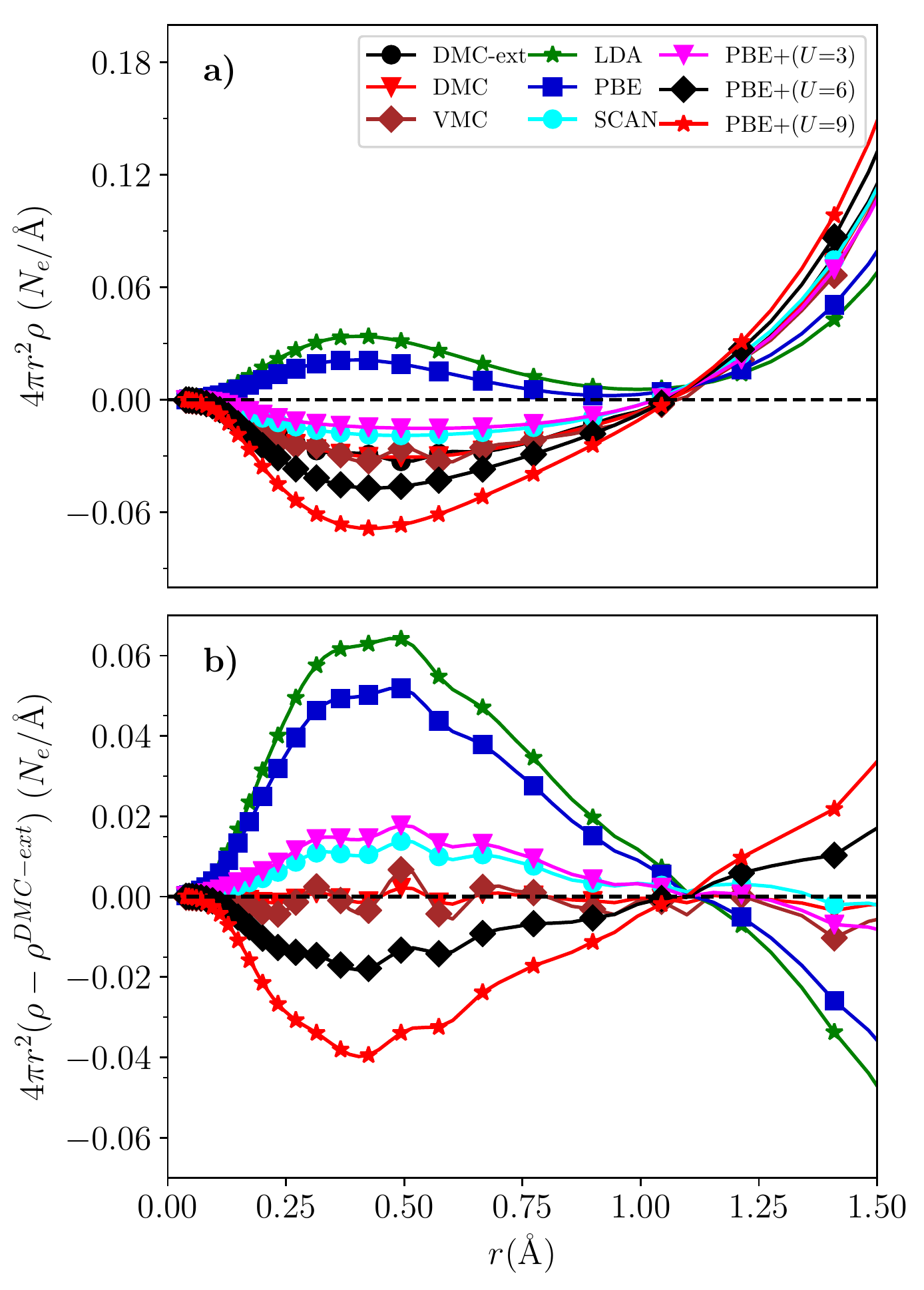}
	\caption{(Color online) a) Radial spin polarization density ($\rho=\rho_{\uparrow}-\rho_{\downarrow}$) and b) radial spin polarization density difference from extrapolated DMC around the O atom with various theoretical methods using RRKJ pseudopotentials. In a),the spin density around O atom is positive for LDA and PBE, while the other theoretical methods yield a negative spin density. DMC spin densities around O in a) and Ni in Fig. \ref{fgr:ni_density} are opposite. Every two out of three markers are omitted for clarity.}
	\label{fgr:o_density}
\end{figure}

The shortcomings of DFT and DFT+$U$ in reproducing experimental redox potentials Li$_x$NiO$_2$ are attributed to the challenges in the accurate description of the hybridization between O-$p$ and Ni-$d$ orbitals \cite{cococcioni2005linear,himmetoglu2014hubbard,shishkin2016self}. Because of this hybridization, it is difficult to correct the exchange-correlation energy term with orbital-dependent energy terms without explicitly accounting for the orbital occupancy of both Ni and O, while taking into account the interatomic coupling. However, the performance of DMC for the voltage curves in Fig. \ref{fgr:volt} suggests that DMC can also provide accurate charge density distributions, and in particular the $p$-$d$ hybridization between Ni and O atoms. 

\begin{figure}
	\includegraphics[width=0.5\textwidth]{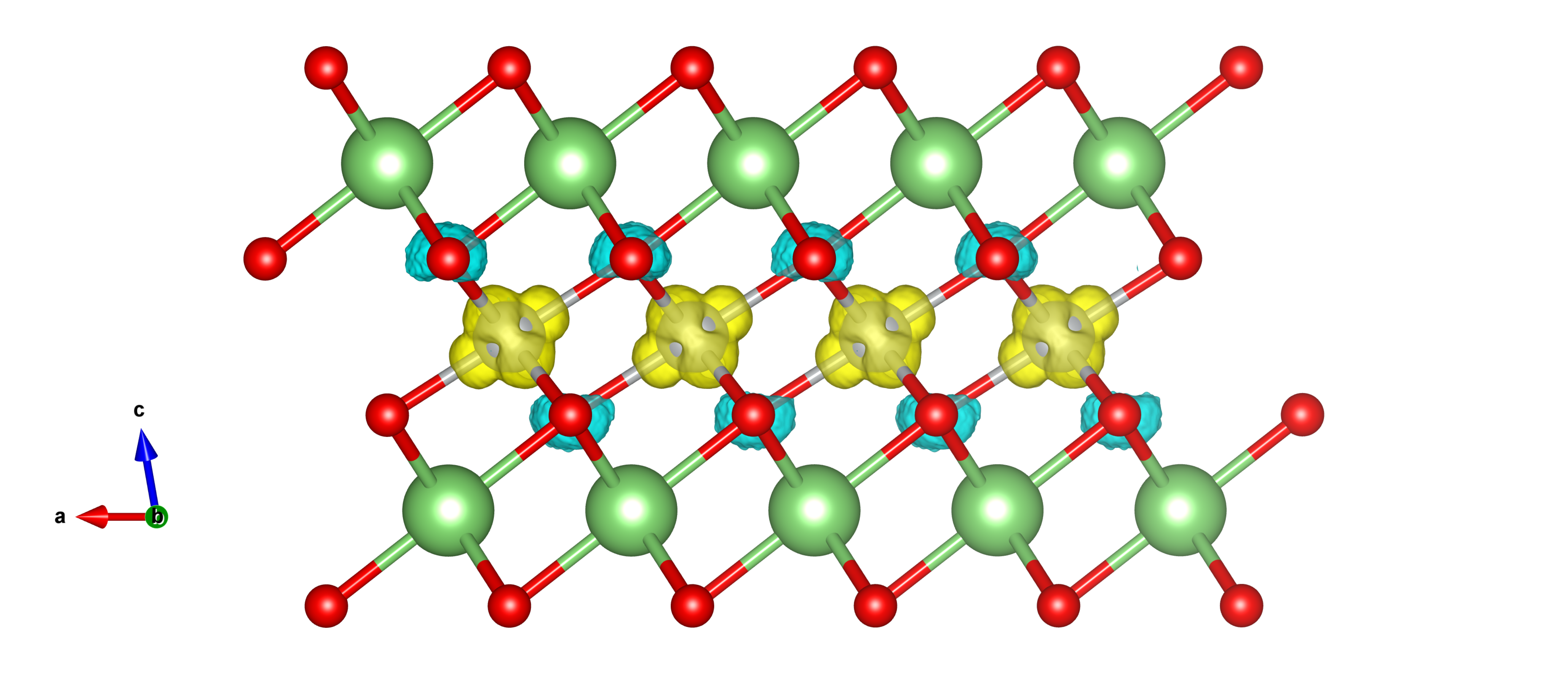}
	\caption{(Color online) LiNiO$_2$ DMC spin density isosurface. Gray, red and green denotes Ni, O and Li atoms. Positive and negative isosurfaces are shown in yellow and blue respectively.}
	\label{fgr:layer_density}
\end{figure}

We investigate the radial spin polarization density of LiNiO$_2$ to understand the degree of hybridization between Ni-$d$ and O-$p$ electrons in Fig. \ref{fgr:ni_density} and \ref{fgr:o_density}. This is motivated by the following: As previously mentioned, in LiNiO$_2$, Ni has a formal charge of 3+, Ni$^{3+}$, while being in an octahedral environment with $t_{2g}^6e_{g}^1$ occupation. This would mean that the $t_{2g}$ manifold is completely occupied. The unpaired electron in the Ni $e_g$ level yields 1 $\mu_B$ magnetization per Ni atom. In LiNiO$_2$, only the Ni-$e_g$ and O-$p$ orbitals have the proper symmetry to hybridize, due to having near 90$^{\circ}$ Ni-O-Ni angles. O-$p$ ($p_x$, $p_y$ or $p_z$) and Ni-$e_g$ orbitals form a filled $e_g$ bonding orbital and a half-filled $e_g*$ anti-bonding orbital \cite{ArroyoydeDompablo2003, y2002first}. 
This is consistent with experimental findings from electron energy loss spectroscopy measurements, which find that Ni$^{3+}$ is in a low spin state in LiNiO$_2$, with a significant hybridization between Ni-$d$ and O-$p$ electrons \cite{koyama2005electronic}. 
Density of states plots in our Supplementary Information also show that the Fermi level is almost purely Ni-$e_g$ and O-$p$ \cite{Supp}. In this respect, the spin polarization density can be used as an indication to show the distribution of the electron at the $e_g*$ level. At the DFT level, the charge density of the $e_g*$ orbital can be obtained through band decomposition of the charge density, but this is not yet accessible within DMC. Therefore, the spread of the spin polarization density can be used to understand if the hybridization is primarily of Ni or O character. Here, we should emphasize that the total spin polarization is constrained at 1 $\mu_B$ per Ni atom in all DFT and DMC calculations in Fig. \ref{fgr:ni_density}a,b and \ref{fgr:o_density}a,b in order to provide a uniform comparison between the methods. Nevertheless, total spin polarization was also found to be close to ~1 $\mu_B$ in DFT calculations where the total spin polarization is completely relaxed \cite{Chakraborty2018}. 

In Fig. \ref{fgr:ni_density}a we show the radial spin polarization density ($\rho=\rho_{\uparrow}-\rho_{\downarrow}$) around the Ni atom in LiNiO$_2$. 
All the theoretical methods in Fig. \ref{fgr:ni_density}a yield a peak density around 0.3 {\AA} and all the values are all positive around the core region of the Ni atom. The peak height depends on the DFT functional, with LDA exhibiting the smallest peak height, followed by PBE, and with other functionals showing larger peak heights. The peak height increases monotonically with the increased +$U$ interaction on the Ni valence electrons. 
To have a closer look at the results of \ref{fgr:ni_density}a, radial spin polarization density differences from extrapolated DMC densities are shown in Fig. \ref{fgr:ni_density}b. 
We highlight three main outcomes from Fig. \ref{fgr:ni_density}b: 
\begin{enumerate*}[label=(\roman*)]
\item DMC and VMC charge densities are almost equal with slight fluctuations around the spin density peak regions, indicating that the trial wavefunction is a good estimate of the true many-body wavefunction 
\item PBE+$U$=6 has the most accurate spin polarization density around Ni atom compared to the extrapolated DMC density. 
This result is illuminating and consistent, as the PBE+$U$=6 calculation also provides the most accurate DFT intercalation voltages in Fig. \ref{fgr:volt}. 
This point will be further discussed below. 
\item PBE+$U$=3 and SCAN functionals provide almost identical densities. 
\end{enumerate*}
Various examples in the literature suggest a reduced self-interaction error in SCAN compared to GGA \cite{Perdew2017, Peng2017, Kitchaev2016, SaiGautam2018a}, which has been reflected in using significantly lower +$U$ values in SCAN compared to GGA+$U$ to produce identical results. 

\begin{figure}
	\includegraphics[width=0.5\textwidth]{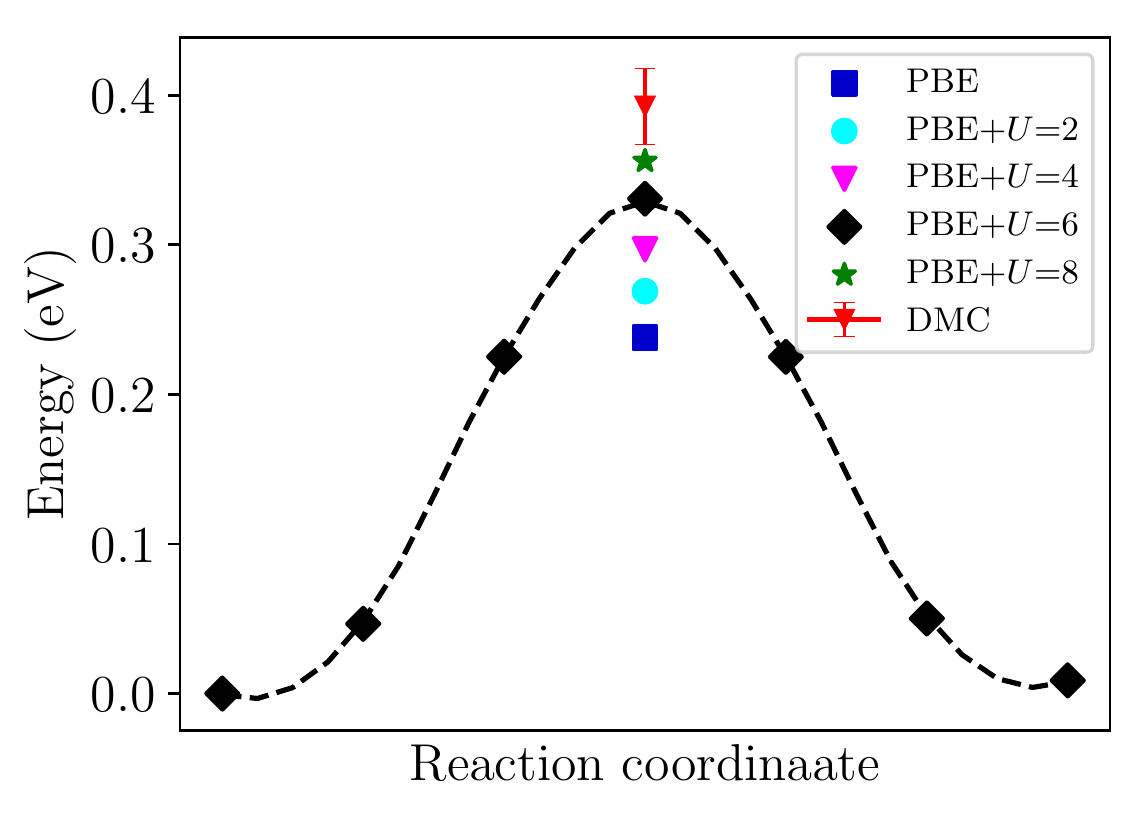}
	\caption{(Color online) Li diffusion barrier in a 14-atom Li$_{0.5}$NiO$_2$ cell. Reaction coordinate, from left to right, indicates the hopping of a Li$^{+}$ ion from the equilibrium site, through the saddle point, Li$_{0.5}$NiO$_2^*$ and to another equivalent equilibrium site. Li$^{+}$ in Li$_{0.5}$NiO$_2^*$ is found in tetrahedral vacancy site as studied in ref. \cite{kang2006factors}. All geometries, including the saddle point are obtained using PBE+$U$=6. Rest of the theoretical methods use these geometries without optimization.}
	\label{fgr:barr}
\end{figure}

In Fig. \ref{fgr:o_density}a and b, we perform the same analysis as in Fig. \ref{fgr:ni_density}a and b, but around the O atom. The LiNiO$_2$ structure we used has the \textit{R-3m} symmetry meaning that all O atoms are identical. This is apparent from the DMC spin polarization density (Fig. \ref{fgr:layer_density}) which shows both $d_{z^2}$ and $d_{x^2-y^2}$ character. Although a Jahn-Teller distortion can ideally be considered, that could lead to splitting between $d_{z^2}$ and $d_{x^2-y^2}$ levels \cite{ArroyoydeDompablo2003, ArroyoydeDompablo2001}. The most important result in Fig. \ref{fgr:o_density}a is that the sign of the spin polarization density changes depending on the DFT functional used. Fig. \ref{fgr:layer_density} shows that the negative spin polarization density on the O atoms are parallel to the Ni-O plane, but not strongly directional on the Ni-O bond axis. A positive spin density around the O atom (e.g. with LDA) in Fig. \ref{fgr:o_density}a is related to a reduced peak value in Fig. \ref{fgr:ni_density}a.  This is correlated to the peak intensities in the density of states (see Supplementary Information), where there is complete overlap between O-$p$ and Ni-$e_g$ peaks at the PBE level, the O-$p$ peak at the Fermi level increases with increasing $U$ value. Local and semi-local DFT leads to delocalized spin polarization densities for the $e_g*$ electron. However, with increasing $U$ interaction and at the meta-GGA level, the $e_g*$ electron is more strongly localized over the Ni core leading to a small polarization over the Ni-O layer. We find that PBE+$U$=3 eV and SCAN functionals produce almost identical spin polarization densities around the O-atom as well, similar to Fig. \ref{fgr:ni_density}a,b. 

\subsection{Lithium diffusion barriers}\label{sec:bar}
In Fig. \ref{fgr:barr}, nudged elastic band (CI-NEB) \cite{jonsson1998nudged} calculations are performed using 5 equidistant images to obtain the Li diffusion saddle point in a 14-atom Li$_{0.5}$NiO$_2$ cell. This cell is found to be sufficiently large to compute the converged energy barrier for lithium diffusion along the tetrahedral vacancy site in LiNiO$_{2}$. The  geometry of the saddle point can heavily depend on the theoretical method used. 
Since the saddle point optimization is not yet available in DMC, we perform several tests to ensure that the saddle point geometry optimized in DFT is reasonable to perform DMC calculations. 

We cross check the saddle point geometries optimized using PBE and PBE+$U$=6 to understand the effect of geometry optimization on the energy barriers. We compare PBE and PBE+$U$=6 as PBE+$U$=6 provides the most accurate voltage curves in Fig. \ref{fgr:volt}. When both of the structures were calculated using PBE, their ground state energies differ by less than 50 meV.  
Our PBE barrier energies compare well with the literature $\sim$ 0.25 eV \cite{kang2006factors}. However, when the same saddle point structures calculated using PBE+$U$=6 eV, their total energies differ by 0.1-0.2 eV. It is known that the distance between NiO$_2$ slabs can drastically effect the diffusion rates \cite{VanDerVen2000}. The reduced slab distance in PBE may lead to increased diffusion barriers in PBE+$U$ calculations. Therefore, we used the saddle point structure optimized using PBE+$U$=6 eV to perform DMC calculations.

Our literature search indicates that a Li diffusion barrier of 0.3-0.6 eV must be expected for Li$_{0.5}$NiO$_2$. Experimental studies on the diffusion rates of LiNiO$_2$ are rather challenging as Ni$^{3+}$ prefers to migrate into Li$^+$ sites at lower Li concentrations. Nevertheless, using $^7$Li NMR spectra a diffusion barrier of 0.6 eV is found \cite{Nakamura2000}. In layered oxides, Li$^+$ diffusion rates are known to increase (diffusion barriers would decrease) with increasing Li-slab distance, hence with decreasing the Li concentration \cite{kang2006factors}. Therefore, the 0.6 eV should be used as an upper bound in our diffusion barrier calculations performed with the Li$_{0.5}$NiO$_2$ cell. Experimental diffusion barrier of LiCoO$_2$ has been studied numerous times with macroscopic \cite{Okubo2009} and local techniques \cite{Balke2012} that yield a diffusion barrier of 0.26-0.3 eV near Li$_{0.5}$CoO2. It has been noted that the diffusion barrier in LiNiO$_2$ should be larger than the diffusion barrier of LiCoO$_2$ \cite{Nakamura2000}. Therefore, the diffusion barrier of Li$_{0.5}$CoO$_2$ can be used as a lower bound for Li$_{0.5}$NiO$_2$. Hence, our analysis yields a range of Li diffusion barrier energies (0.3-0.6 eV) for Li$_{0.5}$NiO$_2$.

Fig. \ref{fgr:barr} shows that the energy barriers are increased with the increased value of $U$, as expected. We find that a $U$ value of 6 eV or larger must be used to obtain Li diffusion barrier energy within 0.3-0.6 eV . DMC Li diffusion barrier energy we calculated, 0.39(3) eV, is larger than all the PBE+$U$ diffusion barrier energies in Fig. \ref{fgr:barr}. 
While PBE+$U$=6 eV reproduces experimental voltage curves, a substantially higher value of $U$, $\sim$10 eV, could be required to fit the energy barrier calculated with DMC.
This result supports previous work showing how the energy of the transition state can depend strongly on the exchange component of the density functional, and often larger exchange ratios are required for accurate barrier height than what is needed for the equilibrium geometries \cite{Lynch2001}. Therefore, it is likely that the $U$ value of 6 eV, which is reasonable for intercalation voltages, leads to an underestimation of the barrier height compared to the DMC barrier height. 

\section{Conclusions}
We showed that it is possible to obtain accurate Li intercalation voltage curves using DMC method, and illustrated this approach on the LiNiO$_2$ layered cathode structure. Semilocal DFT results typically underestimate the voltage curves due to spurious self-interaction effects, and require corrections. However, DMC calculates the electron-electron interactions without any ad-hoc approximations. 
To our knowledge, our work is the first report of a cathode redox potential determined using DMC, which accurately reproduces the experiments and lays the foundation for future methods for predicting redox reaction voltages entirely from first principles. 
We discuss the degree of $p$-$d$ hybridization between Ni and O atoms using spin charge density distributions while comparing to the failures of DFT for this material. We show that charge densities computed by LDA and PBE are significantly different than DMC, while SCAN and DFT+$U$ offer relatively improved results. We show how other material properties such as energy barrier to lithium diffusion can be affected, and how the accuracy of DFT+$U$ may not be transferable across different physical properties.


\bibliography{main}
	
\end{document}